\documentclass[letterpaper]{jpconf}
\usepackage{amsmath, amssymb}
\usepackage{graphicx}
\usepackage[metapost,truebbox]{mfpic}
\usepackage{cite}

\begin{document}

\title{A preliminary threshold model of parasitism in the Cockle\emph{ Cerastoderma edule} using  delayed exchange of stability}

\author{E A O'Grady$^1$, S C Culloty$^2$, T C Kelly$^2$, M J A O'Callaghan$^1$,\\ D Rachinskii$^{1,3}$}

\address{$^1$ Department of Applied Mathematics, University College Cork, Ireland}
\address{$^2$ School of Biology, Earth and Environmental Sciences, University College Cork, Ireland}
\address{$^3$ Department of Mathematical Sciences, The University of Texas at Dallas, Richardson, USA}

\ead{dmitry.rachinskiy@utdallas.edu}

\begin{abstract}
Thresholds occur, and play an important role, in the dynamics of many biological communities.
In this paper, we model a persistence type threshold which has been shown experimentally to exist in
hyperparasitised  flukes in the cockle, a shellfish. Our model consists of a periodically driven slow-fast host-parasite
system of equations for a slow flukes population (host) and a fast {\em Unikaryon} hyperparasite population
(parasite). The model exhibits two branches of the critical curve crossing in a transcritical bifurcation scenario.
We discuss two thresholds due to immediate and delayed exchange of stability effects; and we derive algebraic relationships for parameters of the periodic solution in the limit of the infinite ratio of the time scales.
Flukes, which are the host species in our model,  parasitise cockles and in  turn are hyperparasitised by the microsporidian Unikaryon legeri; the life cycle of flukes includes several life stages and a number of different hosts. That is, the flukes-hyperparasite system in a cockle
is, naturally, part of a larger estuarine ecosystem of interacting species involving parasites, shellfish and birds which prey on shellfish. A population dynamics model which accounts for one system of such multi-species interactions and includes the fluke-hyperparasite model in a cockle as a subsystem is presented. We provide evidence that the threshold effect we observed in the flukes-hyperparasite subsystem remains apparent in the multi-species system. Assuming that flukes damage cockles, and taking into account that the hyperparasite is detrimental to flukes, it is natural to suggest that the hyperparasitism may support the abundance of cockles and, thereby, the persistence of the estuarine ecosystem,
including shellfish and birds. We confirm the possibility of the existence of this scenario in our model, at least partially, by
removing the hyperparasite and demonstrating that this may result in a substantial drop in cockle numbers.
The result indicates a possible significant role for the microparasite in this estuarine ecosystem.
\end{abstract}

\section{Introduction}

Parasites are now widely recognised to play an important role
in the structuring and functioning of ecological communities. In the
estuarine systems of the south western USA, for example, Kuris {\em et al.}
(2008) showed that parasite biomass \textquotedblleft{}exceeded that
of top predators\textquotedblright{} while the biomass of trematodes \textquotedblleft{}equalled\textquotedblright{} that of
the most abundant birds, fish and crustacea. Macroparasites, such as
digenetic trematodes (also commonly known as flukes), have complex
life cycles with several intermediate hosts and complex transmission
dynamics to their final, usually vertebrate, hosts. Although it has
been known for some time that the flukes themselves may be parasitized
by microparasites, the general importance of this relationship in
the population dynamics of the interacting species has only recently
been recognised (Raffel {\em et al.} 2008). The dynamics of many host-parasite
relationships are subject to thresholds of several different types
(Getz and Pickering, 1983; Heesterbeek and Roberts, 1995; Deredec
and Courchamp, 2003; Kalachev \emph{et al.,} 2011). In this section,
we model the impact of a \textquotedblleft{}persistence\textquotedblright{}
type threshold in a hyperparasitised fluke which infects a common
estuarine bivalve mollusc \textendash{} the cockle \emph{Cerastoderma
edule}. A persistence-type threshold is relevant to parasites and
is characterised by a necessary number or density of hosts for parasite
maintenance (\emph{sensu} Deredec and Courchamp, 2003). We begin by
describing in more detail the main species involved in the study.

\subsection{The Cockle}

The cockle\emph{ Cerastoderma edule} populates estuarine regions from
the Barents Sea to Morocco, including the coastlines of Britain and
Ireland, the area in which this study was undertaken. Not only is
this species of commercial importance, but it also plays a significant
role in estuarine ecosystems as an important food source for birds,
fish and crustaceans as well as in the dynamics and stability of the
sediment. Cockle densities at different sites can be highly variable
with $1000$ per square metre not unusual but as many as $54,400$
being recorded (Ducrotoy \emph{et al}, 1991). A range of variables
control cockle population dynamics, including biotic influences such
as bioturbation (that is burrowing activity and aeration of sediment),
predation, parasitism and food availability together with abiotic
factors such as temperature, immersion time, water velocity and sediment
dynamics (Gam {\em et al.}, 2010). In particular, recruitment is significantly
affected by predation by species such as the shrimp \emph{Crangon
crangon} (see for example, Beukema and Dekker, 2005).

\subsection{The Macroparasite - the fluke}

In intertidal ecosystems, flukes are the dominant parasite group (Mouritsen
and Poulin, 2002). These macroparasites exert two influences; firstly
as part of the living diversity, and secondly as diversity indicators,
because their presence is linked to the richness of the free living
fauna (their hosts) as demonstrated by Hechinger\emph{ et al.} (2007),
thereby acting as a proxy for ecosystem health, (Hudson \emph{et al}.
2006). These  digenetic trematodes  generally have
three hosts in their life-cycle: a definitive host where the parasite
reaches sexual maturity, a molluscan first intermediate host in which
asexual reproduction occurs, and a second intermediate host which
provides the vehicle for transfer back to the definitive host (Esch \emph{et
al}., 2002).

\begin{figure}[ht]
\begin{center}
\includegraphics*[width=0.7\columnwidth]{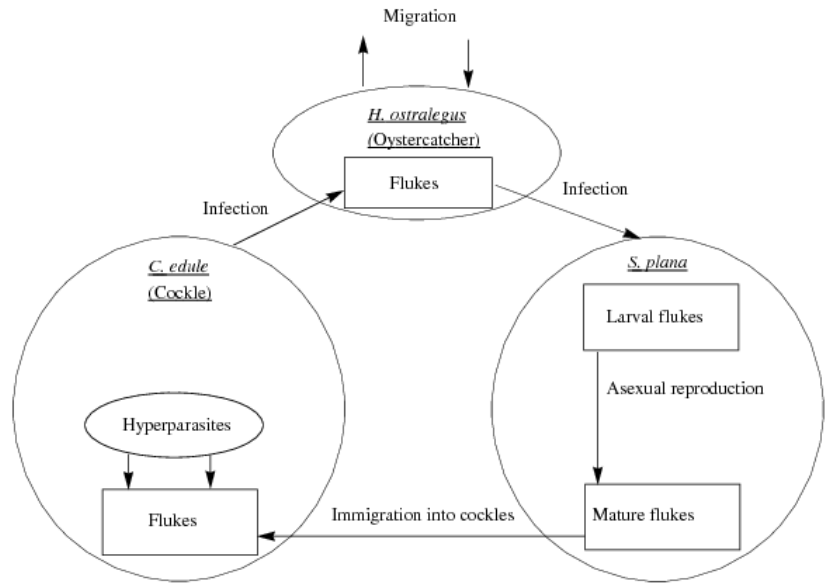}
\end{center}
\caption{Flow chart showing the interactions of species involved and the life-cycle
of the fluke \emph{M. minutus}.\label{fig:Flow-chart-showing}}
\end{figure}

Cockles are first or second intermediate hosts of, at least, 16 fluke
species with complex life cycles involving 2-3 hosts (de Montaudouin
\emph{et al}., 2009; Russell-Pinto {\em et al.}, 2006). This diversity in
parasite infracommunities seems to be possible due to the observed
high degree of spatial segregation of parasite species within their
cockle host (%Lauckner, 1971;
Rohde, 1994). One such fluke is \emph{Meiogymnophallus
minutus},\emph{ }which is one of four sibling species, all of which
infect cockles of the genus\emph{ Cerastoderma}. The fluke can be
found in the mantle epithelium of the cockle (Bowers \emph{et al.},
1996; de Montaudouin \emph{et al}., 2009)\emph{. }The fluke has a body
length of 240-350 \emph{$\mu$m} and a width of about 150 \emph{$\mu$m}
(Russell-Pinto, 1990).\emph{ }This trematode is a member of the Family
Gymnophallidae, with the bivalve mollusc \emph{Scrobicularia plana}
being the first intermediate host, \emph{C.edule }the second intermediate
host, and the oystercatcher \emph{Haematopus ostralegus} - a widely
distributed shorebird - the final, or definitive, host. This cycle
is described schematically in Figure \ref{fig:Flow-chart-showing}.

It has been demonstrated that larval stages of \emph{M. minutus }released
from the first intermediate host and dispersed via water currents,
passively locate their second intermediate host within a distance
of a few hundred metres, where they encyst becoming metacercariae.
These do not reproduce and therefore population growth within the
cockle is solely due to the incremental effects of immigration.\emph{
}The impact of \emph{M. minutus,} when present in large numbers, is
to inhibit the tight closing of the shell valves. Consequently, these
cockles are often found, shell gaping upwards, close to the surface
of the sand (Bowers \emph{et al}., 1996). Cockles infected with flukes
suffer from a range of different effects such as impaired burrowing,
reduced growth, increased mortality and reduced tolerance to opportunistic
microparasitic infections (Lauckner \emph{et al.}, 1983; Wegeberg
and Jensen, 1999; 2003).

In a study conducted along the Northeast Atlantic, this fluke was
found to be the most abundant and widespread trematode in cockles
(de Montaudouin \emph{et al}., 2009) from the Wadden Sea in the north
to Morocco in the south (Gam\emph{ et al.}, 2008). This distributional
range is linked to large scale migrations of shorebirds (Thieltges
and Reise, 2006). In previous studies the prevalence of \emph{M. minutus}
had been found to range from up to 48\% in the Wadden Sea, (Thieltges
and Reise, 2006), to as much as 100\% in Arcachon Bay, France, (de
Montaudouin \emph{et al.,} 2000), in Merja Zerga in Morocco (Gam \emph{et
al}., 2008) and at a number of sites along the south coast of Ireland
(Fermer \emph{et al}., 2011).

\subsection{The Hyperparasite}

Microsporidia are obligate intracellular eukaryotic microparasites,
ranging in size from about $3$ to $6$ \emph{$\mu$m}, most closely
related to the Fungi (Troemel \emph{et al}., 2008). It is clear that
they are up to two orders of magnitude smaller than the fluke. They
infect both vertebrates and invertebrates and include a number of
species that are found in humans (Didier \emph{et al}., 2004). Johnson
\emph{et al}. (2010) describe eight microsporidians, including \emph{Unikaryon
legeri} \textendash{} the subject of this study, that are hyperparasites
of various digenean trematodes which they classify as predators of
the flukes. Although not germane to this study, the issue of whether
\emph{U. legeri }is a parasite (i.e. hyperparasite) or a predator
needs to be addressed by future research because of the implications
for the population dynamics of the interacting species.

\subsection{Interaction between \emph{M. minutus} and \emph{U. legeri}}

Spatial variation in the presence of the hyperparasite, \emph{U. legeri,}
was found by Fermer \emph{et al.} (2010) when cockles ($N=50$ at each
site) were screened at 14 separate locations on the south coast of
Ireland in July 2007. Hyperparasitism occurred at $8$ out of the
$14$ sites and was prevalent only in cockles heavily infected with
the fluke. Of particular significance was the fact that \emph{U. legeri}
was not present where the burden of \emph{M. minutus} was less than
approximately $185$ metaceriae per cockle (Fermer, 2009). This low burden of metacercarial
flukes within the cockle appears to be a threshold enabling infection
with the hyperparasite. It is clearly evident therefore, that the
threshold is that of a persistence type as defined above. It will
ultimately determine the duration of infection, as seasonal effects
will impact strongly on the average numbers of flukes in the cockles
due to their temporal life-cycle.

It has been shown that hyperparasitism results in the death of a high
proportion of the infected flukes. Due to hyperparasite-induced mortality,
the entire metacercarial population has to rebuild every year. Hyperparasitism
is thus a very important factor controlling fluke infrapopulation
size and the overall dynamics of \emph{M. minutus}.

\subsection{Structure of the paper}
In the next section we present and discuss a model of the hyperparasitised flukes in a cockle.
This model is then incorporated as a subsystem in a larger ecosystem model, which is formulated
and tested in Section \ref{big}. The multiscale analysis technique we use is close the method
described by Nizette
\emph{et al}. (2006). The last section contains conclusions.

\section{Fluke-Hyperparasite model}

We present a mathematical model which describes the subsystem occurring
in the individual cockle. In particular, we model the threshold population
of flukes inside a cockle that is needed to provide a sufficient environment
for the parasite population to grow and persist. As explained above,
when fluke populations are less than $185$ per cockle (Fermer, 2009), there are
no hyperparasites present. On exceeding this threshold number of flukes,
there is a significant change in the dynamics of both populations
whereby the hyperparasite population grows, attacks and can kill part
of the fluke population. The parasite has a regulatory effect on the
fluke population which appears to be unique (see Raffel \emph{et al}.,
2008).

\subsection{Threshold model development}

Flukes enter the cockle through immigration as outlined above. We
assume a natural death rate of flukes and an interaction between the
flukes and the hyperparasites that we consider to be of a parasitic
type. On the basis of these assumptions, we write the following equation
for the rate of change of the fluke population:
\begin{quote}
Rate of change of fluke population = (Immigration term) $-$ (Natural
death rate) $-$ (Death due to hyperparasites),
\end{quote}
which leads to the equation:
$$
\frac{dF}{dt}=a(t)-\delta F-\lambda H,\label{dF thresh}
$$
where $F$ is the fluke population, $H$ is the hyperparasite population,
the derivative $dF/dt$ represents the rate of change of fluke population
and $a(t)$ represents a time-dependent immigration rate.\textbf{
}The term $-\lambda H$ represents the death rate of flukes due to
the hyperparasites. As a parasitic type term, we assume that the damage
caused by the hyperparasite to a single fluke is proportional to the
mean number that the fluke carries, that is, $H/F$. Thus the decrease
in the overall fluke population will be proportional to $H$. Here
$\lambda$ is a positive constant representing the number of flukes
lost per hyperparasite per unit time.

For the hyperparasite population, we assume a natural death rate in
the absence of flukes, a positive growth term due to the exploitation
of their hosts, and a compensating term to reflect the damage that
host mortalities have on the population:
\begin{quote}
Rate of Change of parasite population = $-$ (Death rate) $+$ (Growth
term due to parasitism on flukes) $-$ (Compensating term).
\end{quote}
The second, growth, term here was developed as follows. The rate of
production of hyperparasite transmission stages is proportional to
$H$. The magnitude of the proportion of these hyperparasites that
are successful in infecting flukes is characterised by a term reflecting
the relative density of hosts, $F/(F_{0}+F)$, where $F_{0}$ is a
measure of the transmission efficiency, and $F$ is the fluke population,
as described in general terms in Anderson and May (1978). A large
$F_{0}$ indicates that this efficiency is low and is the one that
we use. Under this assumption, that only a small proportion of the
hyperparasite transmission stages actively infect flukes, the growth
term in the hyperparasite population is proportional to $HF$.

The compensating term reflects the fact that as it kills its host,
a certain proportion of hyperparasites are also lost. An important
factor needing consideration here is the distribution of hyperparasites
in the flukes. Consequently, we note that as hyperparasites damage
the fluke population at a rate proportional to $H$, they themselves
will suffer loss proportional to $H$ multiplied by a proportion of
the average distribution $H/F$. We introduce a constant of proportionality
$f$, to indicate this distribution proportion of hyperparasites on
flukes. We will describe this coefficient in greater detail below.
The equation for the rate of change of hyperparasites can then be
written as:

$$
\frac{dH}{dt}=-cH+dHF-\frac{f\lambda H^{2}}{F}.\label{dH thresh}
$$

It is common for systems to have dynamics that interact on differing
timescales (see for example, Etchechoury and Muravchik; 2003) as described
already, and so we now introduce one important assumption to this
last equation. We will require that the interval over which the hyperparasite
population replicates and undergoes transmission occurs on a timescale
much shorter than the corresponding interval of the fluke population.
This is reasonable given the fact that the hyperparasites are several
orders of magnitude smaller than the flukes as mentioned above. To
account for this mathematically, we introduce a new parameter $\varepsilon$
in Eq. (\ref{dH thresh}) and assume $0<\varepsilon\ll1$. Therefore
we obtain the equations:
\begin{eqnarray}
\frac{dF}{dt} & = & a(t)-\delta F-\lambda H,\label{dF thresh final}\\
\varepsilon\frac{dH}{dt} & = & -cH+dHF-\frac{f\lambda H^{2}}{F}.\label{dH thresh final}
\end{eqnarray}

The last term in Eq. \eqref{dH thresh final} can be used to model
the occurrence of over-dispersion of hyperparasites on flukes. Over-dispersion
means that the majority of flukes carry few numbers of hyperparasites,
while a minority are greatly infected. Thus, the death of a fluke
results in a greater than average proportion of hyperparasites being
lost. In this last term, we are concerned with those hyperparasites
living on, and killing flukes, and we reason that the interaction
occurs on timescales of that of the fluke. Thus, when we wish to examine
cases of over-dispersion, it is meaningful to examine the value of
the combination, $f\lambda/\varepsilon$. In particular, values of
this ratio greater than one indicate over-dispersion.

Comparing our model with that described by Anderson and May (1978),
it can be seen that the equation for our host, the fluke, equation
is similar except that we have an immigration term that has a seasonal
dependence instead of a natural birth rate, making the dynamics different.
In the hyperparasite equation, we have an identical death rate and
our birth rate term is comparable in the regime where only a small
fraction of the hyperparasite population is successful in infecting
the flukes as discussed above. The compensating term is equivalent
to Anderson and May's term for host-death induced parasite mortality,
although written here in a different manner. We also have the notable
difference of the differing time scales as introduced through the
small parameter $\varepsilon$. This will be shown to have a major
impact on the dynamics of the system.

\begin{table}
\caption{Parameter values and units used in the model.\label{tab:Parameters-used-in}}
\begin{center}
\begin{tabular}{lll}
\br
Parameter &Description &Value\\
\mr
a&Fluke immigration&20\\
$\delta$&Fluke natural death rate&1/25\\
$\lambda$&Fluke death due to hyperparasitism&1/800\\
$\varepsilon$&Timescale separation parameter&1/5000\\
c&Hyperparasite natural death rate&185 d\\
d&Growth rate of Hyperparasite&1/3500\\
\br
\end{tabular}
\par\end{center}
\end{table}

Table \ref{tab:Parameters-used-in} shows typical parameter values
used in the model. Although direct experimental measurement of parameters
is difficult, we can reason that there is a basis for the order of
magnitude of the values used here. We use a reasonable monthly immigration
rate of $20$ here, particularly to examine the interesting behaviour
of the threshold phenomenon. Natural death rates of both flukes and
hyperparasites, $\delta$ and $c$ respectively, are usually much
less than $1$, and we require that $c>\delta$ due to the fact that
host deaths result in large numbers of parasite deaths. The parameter
$d$ is taken to be small here, to prevent too large a population
of hyperparasites. The fluke death rate due to hyperparasitism, $\lambda$,
is chosen here so that the hyperparasites have an impact on fluke
population and regulate their growth. In the case of over-dispersion
of hyperparasites, we want $f\lambda/\varepsilon$ greater than one
and for our simulations we take a value of $5$. We will see in the
next section that the ratio $c/d$ marks a point where the hyperparasite
population goes to zero, or to a non-zero population, giving immediately
the value used for $c$, if we take this to be the threshold population
and assume it to be $185$ (shown below).

\subsection{Preliminary discussion }

\begin{figure}
\begin{center}
\includegraphics*[width=0.7\columnwidth]{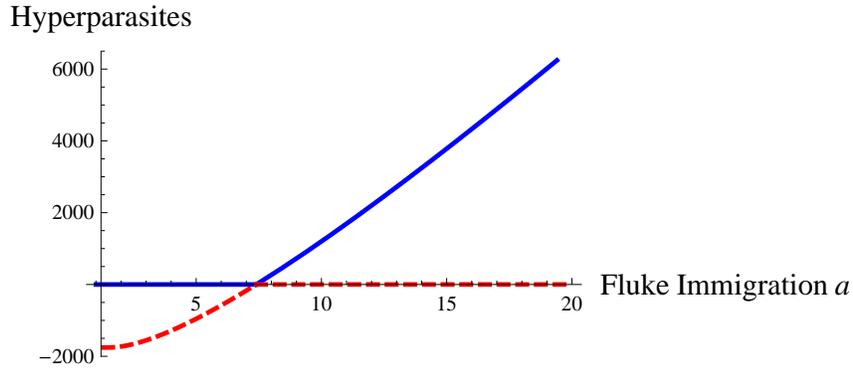}
\par\end{center}
\caption{Bifurcation diagram showing fluke immigration rate parameter $a$
on the horizontal axis with hyperparasite population on the vertical
axis. Transcritical bifurcation point evident at $a=7.4$, where for
this parameter set $(a/\delta)=(c/d)$. Here, the previously stable
zero hyperparasite solution exchanges stability with the previously
unstable non-zero solution. It is evident that for this parameter
set, hyperparasites are absent for $a$ less than $7.4$ and present
above this value. Blue solid lines indicate stable solutions, red-dashed
lines indicate unstable solutions. Here, the parameter values, apart
from the bifurcation parameter $a$, are given in Table \ref{tab:Parameters-used-in}.\label{fig:Bifurcation-diagram for }}
\end{figure}

\begin{figure}
\begin{center}
\includegraphics*[width=0.5\columnwidth]{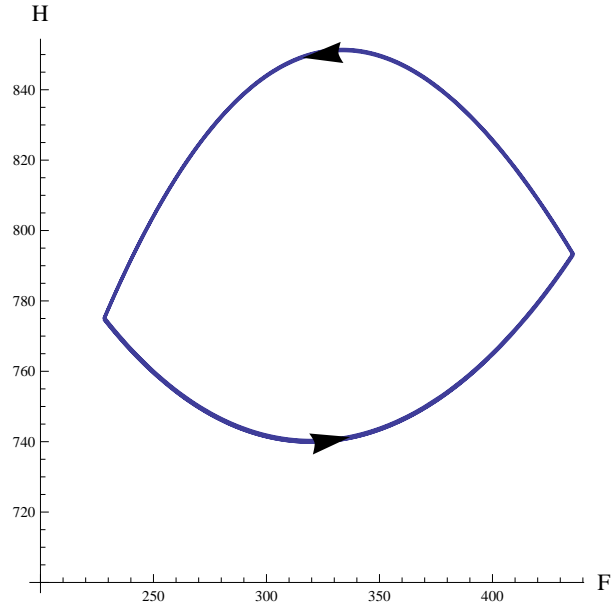}
\par\end{center}
\caption{We include for illustrative purposes the stable periodic regime of
system (\ref{dF thresh final}), (\ref{dH thresh final}) with the
fluke immigration rate $a$ switching periodically between the values
$a=0$ and $a=70$. Here $\varepsilon=1$, i.e. both populations evolve
on the same timescale, $c=185d$, $f=0.9$, $\delta=1/25$, $d=0.00031$ and $\lambda=1/50$. It can be seen that when immigration switches on, there is attraction towards the non-zero equilibrium,
at least after the threshold number of flukes is reached. When immigration
ceases, the system turns back and evolves towards the zero equilibrium,
now stable at this time.\label{fig:Cycles-are-apparent}}
\end{figure}

Two equilibria are possible in the system of Eqs. \eqref{dF thresh final}
and \eqref{dH thresh final} with constant immigration term $a(t)\equiv a$,
namely zero- and non-zero hyperparasite population equilibria, dependent
on the parameters and the threshold effect we wish to model. Taking, $a$,
the constant rate of immigration of flukes into the cockle,
as a bifurcation parameter, a transcritical bifurcation can be shown
to occur at $a=c\delta/d$. This transcritical bifurcation is evident as shown by Figure \ref{fig:Bifurcation-diagram for }. A transcritical bifurcation indicates that a stable positive solution
branches from the zero hyperparasite solution at the bifurcation point.
In the current case, the zero hyperparasite equilibrium becomes unstable
after the transcritical bifurcation while the positive, non-zero,
hyperparasite population becomes stable. Thus, solutions after the
transcritical bifurcation are attracted to the non-zero trajectory,
while solutions before the transcritical bifurcation tend to the stable
zero hyperparasite equilibrium. Therefore, the zero-hyperparasite
equilibrium is stable if the condition $a/\delta<c/d$ is satisfied.
The positive equilibrium exists and is stable if the opposite inequality
holds. This figure is the same for all values of the time scale separation
parameter $\varepsilon$. We see by Table \ref{tab:Parameters-used-in},
that both sides of the inequality have units of flukes. Thus
we recognise that the number of flukes determines the stabilities
of the zero and non-zero hyperparasite populations, and thereby, the
dynamics of the system. In a real-life system, the immigration term
varies temporally due to migratory patterns of birds who consume the
cockles infected with parasites, which ultimately control the cycle
of fluke reproduction, and also due to temperature variation and perhaps
other abiotic influences. If $a(t)$ changes abruptly between two
values, we obtain a switching system forced by a square wave function. Such
a type of function to model the temporal influx of a species has been
used previously in, for example, Holt {\em et al.} (2003). For example,
at certain times of the year, the system can be attracted towards
the zero-hyperparasite equilibrium due to no fluke immigration and
for the rest of the year towards the positive equilibrium, shown in
Figure \ref{fig:Cycles-are-apparent}.

\begin{figure}
\begin{center}
\includegraphics*[width=0.5\columnwidth]{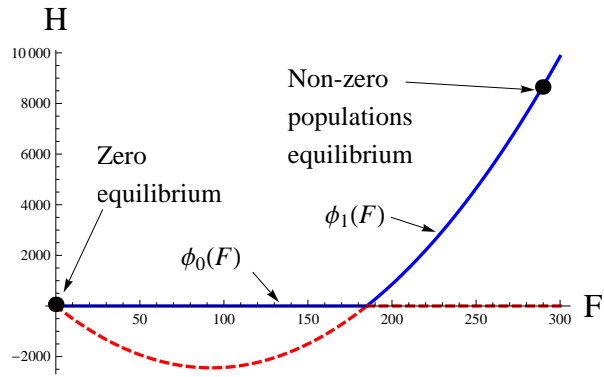}
\par\end{center}
\caption{Intersecting critical curves\textbf{ $H=\phi_{0}(F),\, H=\phi_{1}(F)$
}of system\textbf{ }(\ref{dF thresh final}) and (\ref{dH thresh final})
with parameter values shown in Table \ref{tab:Parameters-used-in}.
Solutions of the dynamical system evolve on these curves, depending
on stability of the equilibria. Two equilibria for the populations
are indicated, and these depend on the value of $a$. When $a=20$,
immigration is switched on and solutions tend to the non-zero equilibrium.
In this case, the zero equilibrium becomes unstable. When immigration
is switched off, $a=0$, the system tends towards the stable zero equilibrium.
\label{fig:PicSlowManifolds}}
\end{figure}

\subsection{Singular perturbation and stability exchange}

First we examine the condition for the bifurcation point evident in
system (\ref{dF thresh final}), (\ref{dH thresh final}).

Taking this system with constant immigration $a$, we wish to examine
the stability of the equilibrium $(F^*,H^{*})=(a/\delta,0)$. The jacobian matrix of
this system is
$$
\left(\begin{array}{cc}
-\delta & -\lambda\\
0 & (-c+dF)/\varepsilon
\end{array}\right).\label{eq:jacobi matrix}
$$
Hence, the eigenvalues are
$$
\mu_{1}  =  -\delta,\qquad \mu_{2}  =  (-c+dF)/ \varepsilon.
$$
For stability of the equilibrium $(F^*,H^{*})=(a/\delta,0)$, we require $\mu_{1},\mu_2<0$
(both eigenvalues are real in this case). Since $\mu_{1}$
is negative by definition ($\delta>0$), the equilibrium is stable
for $c>dF^*=d a/\delta$, with the transcritical bifurcation occurring at
%$F=c/d$.
%Since the corresponding equilibrium is known to be $F^{*}=a/\delta$,
%found by setting Eq. (\ref{dF thresh final}) to zero, we can write
%the relation for the bifurcation point as:
$$
\frac{a}{\delta}=\frac{c}{d}.\label{eq:transcritical bif point}
$$

For $\varepsilon\ll1$, Eqs. (\ref{dF thresh final}) and (\ref{dH thresh final})
become a singularly perturbed system with the hyperparasite being
the fast variable (see above and, for example, Veliov, 1997). Systems
of this kind, where the small parameter $\varepsilon$ tends to zero,
lend themselves to analysis of a reduced initial value
problem, which is one-dimensional in this case. Indeed, the trajectories
of the singularly perturbed system spend most of the time near the
critical manifold, actually a curve in this case, which is defined
by setting the right hand side of the fast hyperparasite equation
to zero, yielding the two roots
$$
H=\phi_{0}(F)\equiv0,\qquad H=\phi_{1}(F)=F(dF-c)/(f\lambda),\label{eq:HH}
$$
(see Figure \ref{fig:PicSlowManifolds}). The critical
curve, which we mention here, indicates a curve that is quickly approached
by solutions which then follow this curve on the $H-F$ plane. The
curve given by $H=\phi_{1}(F)$ is a parabola.
%The fixed points being given by $0$ and $c/d$.
The threshold we
model is the intersection of the two branches of the critical curve,
that is the point $F=c/d$, because the hyperparasite population
is always near zero below this point and is positive, at least sometimes,
above it. Thus, we realise that this results in either positive numbers
of hyperparasites, or zero hyperparasites, depending on the fluke
population and the value of $c/d$, which we take now to be equal
to $185.$ The picture shown in Figure ~\ref{fig:PicSlowManifolds},
can be interpreted either as the phase plane of Eqs. (\ref{dF thresh final})
and (\ref{dH thresh final}), or as the bifurcation diagram of the
fast equation\textbf{ }(\ref{dH thresh final}),\textbf{ }in which
the slow variable\textbf{ $F$} is treated as a parameter. Stable
and unstable equilibrium points of Eq.\textbf{ }(\ref{dH thresh final})\textbf{
}are shown by the solid and dashed lines respectively, with the transcritical
bifurcation at the intersection point where\textbf{ $F=F_{C}=c/d$},\textbf{
}which is the threshold value ($F_{C}=185$).

While a trajectory closely follows one of the curves
\textbf{$H=\phi_{k}(F)$}, the dynamics of the slowly varying fluke
population is approximated by the reduced equation
\begin{equation}
\frac{dF}{dt}=a(t)-\delta F-\lambda\phi_{k}(F).\label{eq:reduced}
\end{equation}
In
the vicinity of the critical curve, solutions evolve on timescale equivalent to
that of the fluke population, that is slow relative to the hyperparasite
population. In our interpretation, the hyperparasite population is
detectable when a trajectory is close to the positive branch \textbf{$H=\phi_{1}(F)$}
of the critical curve, and becomes undetectable when the trajectory
switches to the zero branch \textbf{$H=\phi_{0}(F)\equiv0$}. The
scenario where a trajectory switches from one branch to the other at their intersection point $(F,H)=(F_{C},0)$
is known as an immediate exchange of stabilities, see the descending
part of the trajectory in the left panel of Figure \ref{fig:Seasonal plot}.
It is evident from this figure that a different behaviour also arises
where the trajectory stays in the neighbourhood of the unstable zero
branch for a certain time after passing the point $F=F_{C},\, H=0$.
The trajectory then follows a fast sharp transition from approximately
a point $F=F_{th},\, H=0$, with $F_{th}>F_{C}$, to the stable branch
\textbf{$H=\phi_{1}(F)$}, that is the scenario known as a delayed exchange of stability
(see the ascending part of the trajectory in the left panel of Figure
\ref{fig:Seasonal plot}). It is clear therefore, that our system
exhibits both \emph{immediate} and \emph{delayed} stability exchange,
namely as the populations decrease and increase respectively. The
immediate exchange of stability at $F_{C}$ is defined by the parameters
of the system, while the point of delayed stability exchange at $F_{th}$
is defined by the dynamics of the system and varies for different trajectories. In particular, it is dependent
on how long the value of $a(t)$ is equal to zero and the minimum
value reached by the fluke population, $F_{min}$.

\subsection{Seasonally forced model}

\begin{figure}[h]
\begin{center}
\includegraphics[scale=0.7]{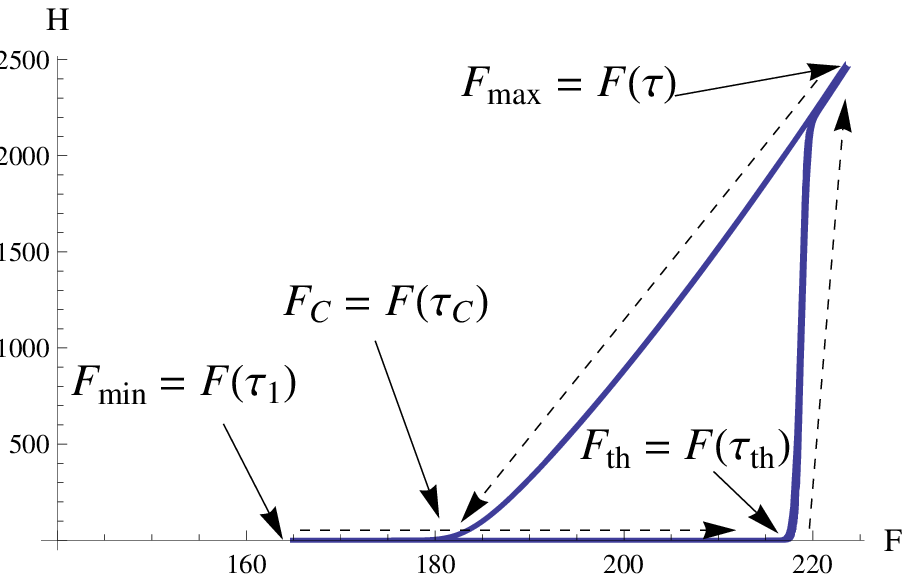}\includegraphics[scale=0.7]{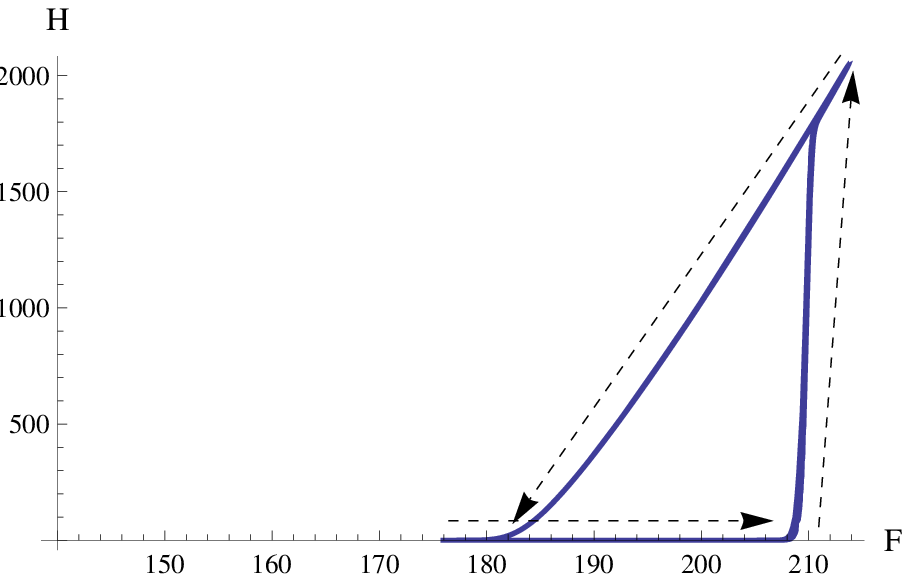}
\par\end{center}
\caption{Left: Periodically varying hyperparasite and fluke populations obtained
by numerical solution of Eqs. \eqref{dF thresh final} and \eqref{dH thresh final}
with $a(t)$ as given by Eq. \eqref{eq:asq}. Parameter values
used were as given in Table \ref{tab:Parameters-used-in}, with $\tau=5$
in Eq. \eqref{eq:asq} and $f\lambda/\varepsilon=5$. Right:
Sinusoidally forced system with fluke dynamics given by Eq. \eqref{eq:dF Sine}.
Here, parameter values used were $a=10$, $c=185d$, $d=1/3000$,
$\lambda=1/300$, $\delta=1/25$, $\varepsilon=1/5000$, $f\lambda/\varepsilon=5$.
Quantitative differences between populations under the two immigration
scenarios are due to different parameter sets e.g. maximum hyperparasite
population is close to $2500$ for square wave immigration term and
just over $2000$ for the sine wave term immigration. \label{fig:Seasonal plot}}
\end{figure}

We see that, in accordance with the different biologies of the interacting
species, the faster timescale of the hyperparasite life cycle results
in two thresholds as shown in the left panel of Figure \ref{fig:Seasonal plot}.
Nevertheless, the transcritical bifurcation point amply illustrates
the overall threshold behaviour of the system.

In the above section we developed a model to describe the hyperparasite-fluke
interaction within a cockle and the presence of a persistence\textendash{}type
threshold. We now address seasonal effects. Seasonality plays
an important role within the system, as effects such as temperature,
bird migration, pollution, and other possible abiotic and biotic factors
as mentioned earlier, will all lead to variations in the populations
under consideration.

As a first approach, we consider Eqs. \eqref{dF thresh final}, \eqref{dH thresh final} with
a square wave, or switch-type immigration term. This means that the
fluke immigration is a positive constant for part of the year, and zero for
the rest. This is a realistic model for our system as flukes mature
within \emph{S. plana}, and are actively moving and swimming after
having been artificially released, only at particular intervals of
the year. In particular, Fermer \emph{et al.} (2010) found that these
stage cercariae appeared from June through October, and were the dominant
developmental stage from July to October. The square wave switching
model is also easier to analyse mathematically than other periodic
patterns of immigration including the sinusoidal type, to be mentioned
later.

We introduce the square wave immigration term $a(t)\equiv a(t+T)$ in Eq. \eqref{dF thresh final}
where $a(t)=A>0$ for part of the year and zero otherwise, such as
\begin{equation}
a(t)=\begin{cases}
A, & 0\leqslant t\,\mbox{(mod \ensuremath{T})}\leqslant\tau\\
0, & \tau<t\,\mbox{(mod \ensuremath{T})}<T=12\end{cases}\label{eq:asq}
\end{equation}
where there is immigration over $\tau$ months, and the population
suffers from natural death and hyperparasitism as before. As an initial
approximation, we take the value of $\tau$ to be 5, meaning we allow
immigration for five months in the year, in keeping with experimental
findings described above. A number of important population points
and times occur and are shown in the left hand panel of Figure \ref{fig:Seasonal plot}.

\begin{figure}[h]
\begin{centering}
\includegraphics[scale=1.1]{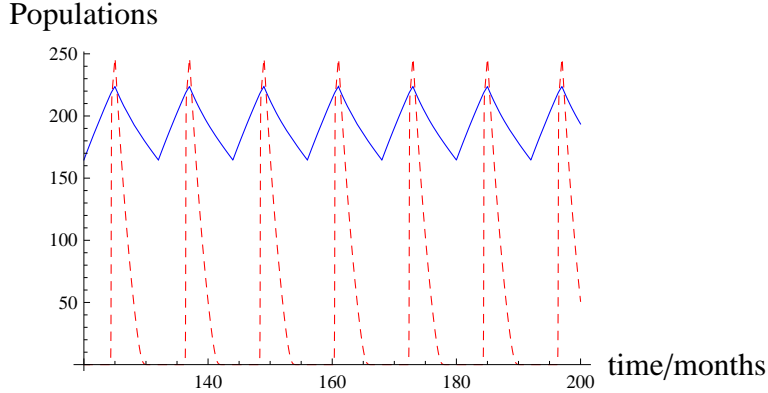}
\par\end{centering}
\caption{Time trace of the system given by (\ref{dF thresh final}), (\ref{dH thresh final})
and \eqref{eq:asq} showing the stable periodic dynamics of the
fluke population in blue, hyperparasite population (factor of $1/10$)
in dashed red, simulated after truncation of the initial variation. Parameters
used are as given in Table \ref{tab:Parameters-used-in} with $H_{0}=F_{0}=100$
at $t=0$, $\tau=5$ in Eq. \eqref{eq:asq} and $f\lambda/\varepsilon=5$.
Yearly periodic behaviour is evident for both populations.\label{fig:long periodic soln}}
\end{figure}

This figure presents the trajectory of a stable $T$-periodic solution
of system \eqref{dF thresh final}, \eqref{dH thresh final} and \eqref{eq:asq}.
Figure \ref{fig:long periodic soln} presents the time traces of both
of the populations (for clarity, the hyperparasite population is scaled
by a constant) over a longer simulation, demonstrating the stability
of the system and the presence and absence of the hyperparasite population
depending on the oscillation of the fluke population above and below
the threshold. In the left hand panel of Figure \ref{fig:Seasonal plot},
the fluke population reaches its minimal value $F_{min}$ at the time of the year
when the immigration starts, that is at the moment
$\tau_{1}=0$ (and then at the same time of the year $t=k T$ every year). At this point, there are no hyperparasites
present in the system (more precisely, hyperparasites are not detectable
as the population $H$ is close to zero and in effect becomes occult).
As the fluke population increases with time it eventually reaches
a value $F_{th}>F_{C}$ at a moment $\tau_{th}$, whereby the delayed
loss of stability occurs and the system quickly tends to the stable
positive branch, $H=\phi_{1}(F)$ of the critical curve. Further increase
in both populations continues, until the instant $\tau$ whereupon
the fluke population tends to die off through mortality caused by
the hyperparasites, together with zero immigration and natural death.
The fluke and hyperparasite populations decrease along the positive
branch $H=\phi_{1}(F)$ towards the point $(F,H)=(F_{C},0)$. The hyperparasite
population has now collapsed to an exceedingly low level (i.e., becomes
undetectable) after the moment $\tau_{C}$ when the fluke population
reaches the value $F_{C}$. The fluke population continues decreasing
to the minimal value $F_{min}$, which is achieved at the moment $T$,
whereupon the new period begins. As we take monthly time intervals,
$T=12$ in this case.

We see that in the limit of the time separation parameter $\varepsilon$
tending to zero, that is, the populations evolve along the critical
curve, as indicated in the left panel of Figure \ref{fig:Seasonal plot},
the periodic solution is characterised by a few parameters, namely
the population values $F_{min}=F(0),\, F_{th}=F(\tau_{th}),\, F_{max}=F(\tau)$
and the time moments $\tau_{th},\,\tau,\,\tau_{C}$. In
this limit, the periodic solution follows the zero branch, $H=0$,
between the moments $\tau_{1}=0$ and $\tau_{th}$ and the positive
parabolic branch, $H=\phi_{1}(F)$, on the rest of the period, while
the transition from the zero branch to the positive branch becomes
infinitely fast. It is interesting in this regime to look further
at the algebraic consequences due to the ability to integrate the
system explicitly.

\subsection{Algebraic relations for the limit of the periodic solution\label{sec:Derivation-of-Algebraic}}

The reduced equation \eqref{eq:reduced} can be integrated explicitly
on each branch of the slow manifold. On the time interval $0\leqslant t\leqslant\tau_{th}$,
the reduced equation is obtained by setting $H=0$ and $a(t)=A$,
resulting in a linear equation with the solution
\begin{equation}
F(t)=F_{A}(t)=\frac{A}{\delta}+e^{-\delta t}\left(F_{min}-\frac{A}{\delta}\right).\label{eq:F tilde square}
\end{equation}
We can immediately deduce the value for one characteristic
population,
$$
F_{th}=F_{A}(\tau_{th})=\frac{A}{\delta}+e^{-\delta\tau_{th}}\left(F_{min}-\frac{A}{\delta}\right).\label{eq:rel 1}
$$

Between the moments $\tau_{th}$ and $\tau_{C}$, when parasites are
present after the threshold time and the solution follows the positive
branch $H=\phi_{1}(F)$, the evolution of the fluke population is
governed by the reduced equation:
$$
\frac{dF}{dt}=\alpha-\delta F-\frac{F}{f}\left(-c+dF\right)\label{eq:df/dt with paras}
$$
where $\alpha=A$ on the time interval $\tau_{th}\leqslant t\leqslant\tau$
and $\alpha=0$ on the time interval $\tau\leqslant t\leqslant\tau_{C}$.
Separating the variables in this equation and integrating
over each of these two time intervals leads to the relations
$$
\phi_{A}(F_{max})-\phi_{A}(F_{th})=\tau-\tau_{th},\qquad\phi_{0}\left(c/d\right)-\phi_{0}(F_{max})=\tau_{C}-\tau,\label{eq:relation 1}
$$
with
$$
\phi_{\alpha}(F)=\int\frac{dF}{\mu F^{2}+\nu F+\alpha},\label{eq:Phi(F) integral}
$$
where $\mu=-d/f$, $\nu=c/f-\delta$. As the discriminant $\nu^{2}-4\,\mu\alpha$
of the denominator of the integrand
is positive (we neglect the special case where $c/f=\delta$),
$$
\phi_{A}(F)=\frac{1}{\sigma}\log\Big|\frac{2\mu F+\nu-\sigma}{2\mu F+\nu+\sigma}\Big|,\qquad\phi_{0}(F)=\frac{1}{\nu}\log\Big|\frac{2\mu F}{2\mu F+2\nu}\Big|,\label{eq:solution to integral}
$$
where $\sigma=\sqrt{\nu^{2}-4\mu A}$.

From the moment $t=\tau_{C}$ to $t=T$ the solution again follows
the zero branch $H=0$ of the slow manifold, like during the time
interval $0\leqslant t\leqslant\tau_{th}$ considered above. However,
now $a(t)=0$ and the reduced equation yields
\begin{equation}
F(t)=F_{0}(t)=\frac{c}{d}e^{\delta\left(\tau_{C}-t\right)}\label{eq:F0}
\end{equation}
which gives a result for the minimum fluke population reached, namely,
$$
F_{min}=\frac{c}{d}e^{\delta(\tau_{C}-T)},\label{eq:fmin, T}
$$
at the time $t=T.$

Finally, consider the time interval when the parasite population is
small, i.e. the trajectory stays near the zero branch of the critical curve from the moment $\tau_{C}$ to the moment $\tau_{th}+T$. The delayed loss of stability phenomenon ensures that, in the limit
of vanishing $\varepsilon$, the initial and the final moments of
this time interval are related by the integral equation
\begin{equation}
\intop_{\tau_{C}}^{\tau_{th}+T}\left(-c+F(t) d\right)dt=0, \label{eq:integral eqn square wave}
\end{equation}
see for example Butuzov {\em et al.} (2004).
This relation follows from the explicit solution of the linearised
Eq. \eqref{dH thresh final} at $H=0$,
$$
\varepsilon\frac{dH}{dt}  =  (-c+dF)H;
$$
the solution reads
$$
H(t)=H(t_0)\left({\rm Exp}\left[\frac{1}{\varepsilon}\int_{t_{0}}^{t}(-c+F(t)\, d)\, dt\right]\right)\label{eq:H(t) integral eq-1}
$$
with a large parameter $1/\varepsilon$ under the exponent.
%where we know that $F(t)$ is given by Eq. \eqref{eq:F tilde square}.
We see that $H(t) = H(t_0)$ if the integral under the exponent is zero,
which leads to the relationship \eqref{eq:integral eqn square wave} between the moments $\tau_C$ and $\tau_{th}+T$ in our model.

Using the periodicity of the solution and substituting the expressions
(\ref{eq:F tilde square}), (\ref{eq:F0}) for $F$ on the time intervals
$0\le t\le\tau_{th}$, $\tau_{C}\le t\le T$ in Eq. \eqref{eq:integral eqn square wave},
we obtain
$$
c\tau_{C}-cT-\frac{c}{\delta}e^{\delta(\tau_{C}-T)}+\frac{c}{\delta}-c\tau_{th}+\frac{dA\tau_{th}}{\delta}+\left(\frac{1}{\delta}-\frac{e^{-\delta\tau_{th}}}{\delta}\right)\left(d\, F_{min}-\frac{dA}{\delta}\right)=0.\label{eq:relation eq. long}
$$

In summation, we have obtained five equations with five unknown
parameters, namely $F_{min},$ $F_{max},$ $F_{th},$ $\tau_{C},$
$\tau_{th},$ of the periodic solution, in the limit of small $\varepsilon$:
\begin{equation}
\begin{array}{c}
F_{th}=\frac{A}{\delta}+e^{-\delta\tau_{th}}\left(F_{min}-\frac{A}{\delta}\right), \\
\phi_{A}(F_{max})-\phi_{A}(F_{th})=\tau-\tau_{th}, \\
\phi_{0}\left(\frac{c}{d}\right)-\phi_{0}(F_{max})=\tau_{C}-\tau, \\
F_{min}=\frac{c}{d}e^{\delta(\tau_{C}-T)}, \\
c\tau_{C}-cT-\frac{c}{\delta}e^{\delta(\tau_{C}-T)}+\frac{c}{\delta}-c\tau_{th}+\frac{dA\tau_{th}}{\delta}+\left(\frac{1}{\delta}-\frac{e^{-\delta\tau_{th}}}{\delta}\right)\left(d\, F_{min}-\frac{dA}{\delta}\right)=0. \end{array}\label{eq:Set of relational eqns (a)}
\end{equation}

\subsection{Numerical results\label{sub:Numerical-results}}

The set of Eqs. \eqref{eq:Set of relational eqns (a)}, can be rewritten
to yield two equations with two unknowns,
$\psi_{1}(\tau_{th},\tau_{C})=0$ and $\psi_{2}(\tau_{th},\tau_{C})=0$.
These equations are not given explicitly here as their form does not contribute
any additional information to understanding of the problem.

\begin{figure}[h]
\begin{center}
\includegraphics[scale=0.8]{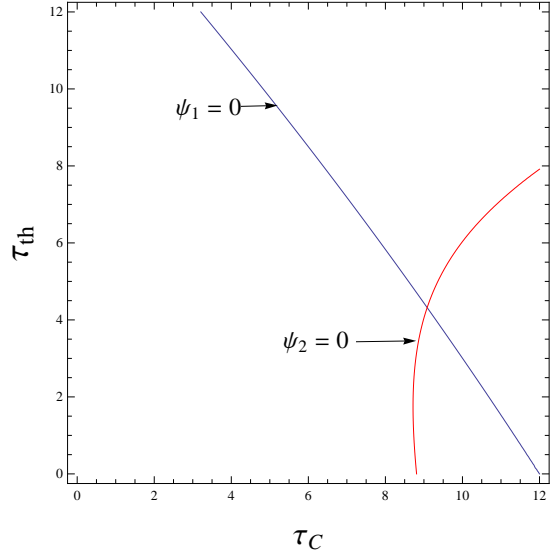}
\par\end{center}
\caption{Plot showing intersection of the functions $\psi_{1}=0$ and $\psi_{2}=0$
for estimating $\tau_{th}$ and $\tau_{C}$. Parameters are as given
in Table \ref{tab:Parameters-used-in}, with $\tau=5$ in Eq. \eqref{eq:asq}
and $f\lambda/\varepsilon=5$.\label{fig:Contour-plot-showing}}
\end{figure}

Using the known set of parameter values given in Table \ref{tab:Parameters-used-in},
a plot of both functions gives an estimate for a point of intersection.
This estimate is then used to numerically calculate the point of intersection
giving the actual values of $\tau_{th}$ and $\tau_{C}$. For the
parameter set used in Table \ref{tab:Parameters-used-in} and $\tau=5$
in Eq. \eqref{eq:asq}, the graphical output is shown in Figure
\ref{fig:Contour-plot-showing}. The actual solution found using the
estimates from the plot yields $\tau_{th}=4.3$ months and $\tau_{C}=9.1$
months.

Using the results for $\tau_{th}$ and $\tau_{C}$, we obtain
other parameters from the relational equations. That is, we find that
\begin{center}
$F_{th}=218$, $F_{min}=165$, and $F_{max}=223$.
\par\end{center}

\begin{figure}[h]
\begin{centering}
\includegraphics[scale=0.9]{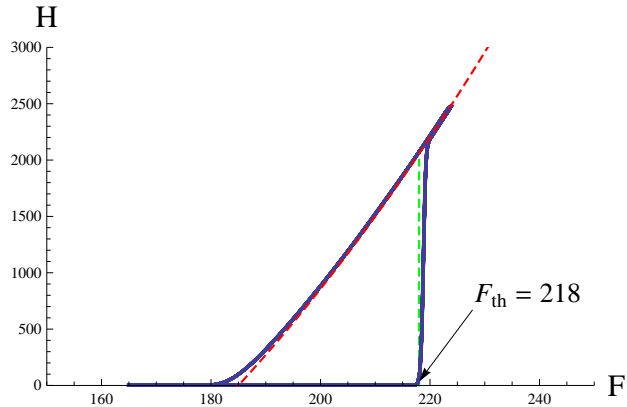}
\par\end{centering}
\caption{Numerical simulation of the parasite population as a function of fluke
population (solid) showing the critical curve and infinitely
fast transition (dashed) at the point $F_{th}=218$ obtained in the limit $\varepsilon\to0$ from algebraic Eqs. \eqref{eq:Set of relational eqns (a)} for the parameter set in Table
\ref{tab:Parameters-used-in}.\label{fig:Numerical manifolds}}
\end{figure}

Using the algebraic relations to calculate results as above, it
is interesting to compare the numerical solutions with the theoretical
infinitely fast transition between both branches in the limit $\varepsilon\rightarrow0$.
Figure \ref{fig:Numerical manifolds} indicates this with the threshold
point $F_{th}=218$ calculated from above. A close
correlation between numerics and the limit solution obtained from Eqs. \eqref{eq:Set of relational eqns (a)} is evident.

\subsection{Sinusoidal immigration term}

To confirm the applicability of the model, another variation to the
immigration of flukes into the cockle is considered, namely that of
using a sinusoidal type immigration, as mentioned already. We state
this new modification by changing the immigration term in Eq. \eqref{dF thresh final}
to
$$
\frac{dF}{dt}=a\,(1+\sin(2\pi t/T))-\delta F-\lambda H,\label{eq:dF Sine}
$$
where the immigration rate is periodic with the period $T$ of one
year. Immigration begins again at time $\tau_{1}=0$ where the fluke
population is at its minimum $F_{min}$. The dynamics follow the same
pattern as described above, shown in the right hand panel of Figure
\ref{fig:Seasonal plot} demonstrating that the qualitative dynamics
of the model remain unchanged. However, this type of model does not give us explicit algebraic relations for parameters of the
periodic solution in the limit $\varepsilon\to0$.

\section{Bird-Shellfish-Parasite model}\label{big}

In the previous section, we have considered a model demonstrating the threshold
effect observed for the hyperparasite population to grow and persist
in a single cockle, depending on the number of flukes present.
We now extend the model to include other species that have
an impact on flukes-hyperparasite subsystem, see Figure \ref{fig:Flow-chart-showing}. This can help understanding
of the importance that the hyperparasite plays, not just to cockle
health, but to the entire ecosystem in bay areas.
%The reason that
%we need to look further at incorporating other species is clear from
%the understanding we obtained from the previous model. We initially
%concentrated on a square-wave function for fluke immigration. We saw
%that this periodically forced the hyperparasite towards either the
%zero or non-zero equilibrium solutions. We showed subsequently that
%changing this to a more general sinusoidal immigration term left the
%interesting dynamics of the threshold phenomenon intact. However,
The periodic immigration terms we used in the flukes equation are artificial, as
the actual immigration is derived from the seasonal migration
of birds to and from the area. Therefore, we attempt to model the threshold
phenomenon using more realistic dynamics related to both the carriers of
the initial infection, that is the birds, and the platform for mass
reproduction, that is the mollusc \emph{S. plana}. The aim of this
section will be to describe each population
individually but yet maintain the threshold of an average number of
185 flukes per cockle necessary for the hyperparasite population to
grow and persist.

\subsection{Oystercatcher population}

It can be seen from Figure \ref{fig:Flow-chart-showing} that the
driving force behind the dynamics of the fluke life-cycle is the
Oystercatcher (\emph{Haematopus ostralegus}) population. These are
the definitive host of the fluke and are the link between the fluke
in cockle and those entering \emph{S. plana, }so as to provide a
new cycle of immigration and infection. That is, they play the necessary
part of closing the life-cycle of the fluke. We refer to these
Oystercatchers simply as birds. These birds are also
the main predators of the cockle population. Therefore, it is beneficial
for the fluke to have birds eat as many cockles as possible, to result
in high levels of infection. A break in this point of the chain would
mean no flukes present and in turn, no hyperparasites, once the lower
limit of the threshold is reached. The fact that the birds consume
huge numbers of cockles is itself of interest to the system. Questions
as to the lower numbers of cockles needed to entice birds to an area
can be asked.

\begin{figure}[h]
\begin{centering}
\includegraphics[scale=0.8]{%Plot_of_bird_forcing_term.eps
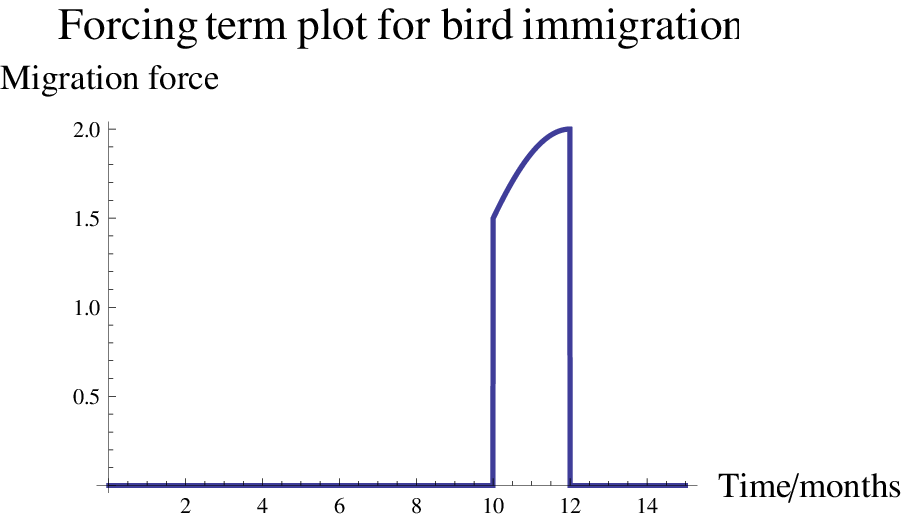} \ \includegraphics[scale=0.8]{%Plot_of_mature_flukes_forcing_term.eps
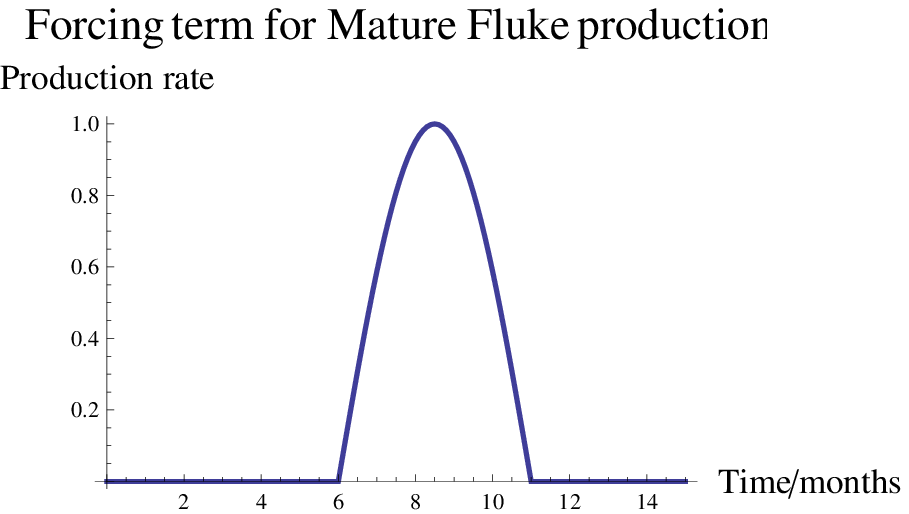}
\par\end{centering}
\caption{Left: The bird migration forcing term $p(t)$ showing immigration beginning
in October, continuing at a declining rate until the end of December,
and birds leaving the area in January. This is periodic on a 12 month
interval. Right: Plot showing the forcing function $\kappa(t)$ for mature fluke production.
Mature flukes were found to be actively swimming and capable of infecting
cockles from June (month 6) to November (month 11).\label{fig:-piecewise-bird forcing}}
\end{figure}

Since these birds are migratory, we introduce a periodic forcing
term. We allow the bird population $B$ to follow a basic logistic
equation, modified however to be dependent on the cockle population $C$.
This ensures that low numbers of cockles lead to birds having less
interest in returning to, or coming to, the area. It is, however,
evident that they have an ability to prey on food sources other
than that of cockles, but the cockle does provide the main source
of food for these birds and thus provides the major stimulus for bringing
birds to the area. Thus, the equation for the bird population $B$
will be given by
$$
\frac{dB}{dt}=k p(t) CB+k_{0}B-sB^{2},\label{eq:db/dt}
$$
where the migration $p(t)$ is periodic with 12 month cycles; $k$ is a scaling coefficient. It is observed that
birds migrate to area's of Ireland in large numbers for Winter months,
with peak numbers seen between September and March (Birdwatch Ireland.
http://www.birdwatchireland.ie/Default.aspx?tabid=314; accessed 8th
March 2012). This forcing term is adjusted to allow for these dates
to influence the migratory patterns. As an approximation, we set up
the forcing function $p(t)$ so as bird immigration begins on the
$10^{th}$ month, that is October, and all birds leave at the beginning
of January; in the interim, there is an added period of growth in
bird population, see the left panel of Figure \ref{fig:-piecewise-bird forcing}.

The growth in logistic form due to cockles is accounted for by the
$CB$ of the first term and the competition between birds, by
the last term. The second term allows for a growth in the bird population
even with low cockle numbers indicating alternative food sources.
It should be noted that we assume all birds are carriers of the fluke,
or at least pick up the fluke infection immediately on returning to
an area with infected cockles, and so all of the population can spread
flukes to \emph{S. plana}. This assumption is supported by the
fact that most areas with fluke infections show cockles with a prevalence
of over 80\% and reaching 100\% (Fermer \emph{et al}., 2010) and the
observation that the birds are each capable of eating in excess of
500 cockles per day. %(Davidson, 1968).

\subsection{{\em Scrobicularia plana} populations}

We refer to \emph{S. plana} just as plana, and we make
an initial assumption here that it is reasonable to divide the population
into an overall constant population $P_{0}$, and the subgroup of
fluke-infected plana, $P_{I}=P_{I}(t)$.
The infected plana population grows due to susceptible plana $P_{0}-P_{I}(t)$
encountering flukes in an infectable stage, that is, those deposited
by birds. The population decreases due to a recovery term whereby
infected plana recover to carry no flukes. Thus we assume growth
in the infected population proportional to the product of the susceptible
plana and bird populations, and decline proportional to infected plana,
that is standard infection-recovery terms:
$$
\frac{dP_{I}}{dt}=\xi B(P_{0}-P_{I})-\gamma P_{I}.\label{eq:dpi/dt}
$$

According to Fermer \emph{et al}. (2010), the prevalence of fluke infection
in plana ranged from over 10\% to nearly 25\% over the course of a
year. The numbers of plana in an area was also found to be much less
than those of cockles. Due to the possible high densities of cockles,
we assume cockle populations in an area to be of the order of millions.
%The dynamics of the cockle population will be addressed below.
It seems a reasonable assumption that an overall plana population of
tens of thousands in an area is acceptable. We use the value of $P_{0}=20000$
and obtain %hope to have
the population of infected plana varying from about
$2000$ to $5000$.

\subsection{Fluke populations}
The mechanism of flukes progressing through the system is complex
and needs to be compartmentalised so as to allow study of individual
sections and effects. The model we have chosen is to have a first
stage of plana infection, what we call larval flukes, $F_{PL}$. These
are those flukes that initially infect plana from birds. The population
will be proportional to the infected plana population ($F_{PL}=\beta P_{I}$),
and we assume small numbers on average per plana. These larval flukes
undergo asexual reproduction within the fluke to produce large numbers
of mature flukes, $F_{PM}$, still inhabiting plana. These mature
flukes develop only at certain times of the year, so we
introduce a periodic forcing term to this equation. The mature flukes
go on to infect cockles through emigration from plana and meeting
a cockle, but will undergo a death rate either within the plana or
during the transmission process, hence we add a natural death rate term
to this equation leaving
$$
%\begin{equation}
\frac{dF_{PM}}{dt}=r \kappa(t)F_{PL}-\epsilon F_{PM},\label{eq:dFpm/dt}
%\end{equation}
$$
where the plot of $\kappa(t)$ is shown in right panel of Figure \ref{fig:-piecewise-bird forcing}; $r$ is a scaling coefficient. These completely
developed and freely swimming flukes were observed between June and
November, and were the dominant stage from July to October (Fermer
\emph{et al}., 2010).
%Similarly to the periodic bird migratory term
%described above, we use a piecewise periodic function to force this
%stage fluke to appear from June to November and show this piecewise
%plot for a period of one year in Figure .
%

The next fluke population, $F_{CO}$, is that proportion of mature flukes that successfully
infect a cockle. The number of flukes in cockles should not exceed the number
of mature flukes as they have no means of reproducing once inside
cockles.
 The population $F_{CO}$ of flukes in cockles
%(not to be confused with the fluke threshold population also defined as $F_{CO}$ above),
grows proportionally to the product of the mature fluke population
and the cockle population, as per typical infection rate term. They
have a natural death rate within the cockle and, importantly,
suffer due to parasitism by the hyperparasite.
%The equation may be written as
%$$
%\frac{dF_{CO}}{dt}=f\, C\, F_{PM}-\delta F_{CO}-\nu H.
%$$
Furthermore, we include an
additional  term in the equation for $F_{CO}$. As flukes
damage cockles, it is reasonable to assume that they in turn suffer
loss as this damage increases. We assume that flukes
damage cockles as a parasite giving a term of the form $-\omega F_{CO}$
in the cockle equation, yet to be derived. Therefore, flukes themselves
will be lost with a compensating term proportional to $-\omega F_{CO}\times(F_{CO}/C)$.
We give the full equation as
%\begin{equation}
$$
\frac{dF_{CO}}{dt}=f\, C\, F_{PM}-\delta F_{CO}-\nu H-\frac{\zeta\omega F_{CO}^{2}}{C},\label{eq: dFc final}
$$
%\end{equation}
where the parameter $\zeta$ represents the possibility for over-dispersion
as in Eq. (\ref{dH thresh final}). Comparing this equation to Eq. (\ref{dF thresh final}),
the artificial immigration term $a(t)$ is replaced by a dependence
on both the cockle population, $C(t)$, and the mature fluke population, $F_{PM}(t)$.
The cockle population varies throughout the year due to
recruitment and reproduction, as well as predation by birds during Winter months.
The second and third terms of both equations
remain equivalent. The reason for the additional final term compared
to Eq. (\ref{dF thresh final}) is because system (\ref{dF thresh final}),
(\ref{dH thresh final}) did not include a measure of cockle damage
due to flukes as their population was not relevant to the model.

\subsection{Cockle population}

The cockle population is assumed to grow logistically in the absence
of outside negative effects. It declines due to predation by birds
(proportional to the product of cockle and bird populations) and due
to fluke damage (proportional to the population of flukes in cockles). The equation is
%\begin{equation}
$$
\frac{dC}{dt}=\lambda C-\mu C^{2}-\omega F_{CO}-hCB.\label{eq:dc/dt}
$$
%\end{equation}
Parameters $\omega$ and $h$ represent the negative parasitic
effect of flukes and the predatory effect of birds, respectively.

\subsection{Hyperparasite population}

The final species we need to model is the hyperparasite. This
retains a threshold type behaviour whereby when the population of flukes
drops below $185$ on average per cockle, the hyperparasite is undetectable.
The hyperparasite has a natural death rate, and also a
compensating term due its impact on the fluke population, similar
to that described earlier for the fluke population and its effect
on cockles. The growth rate in this case is reasoned as follows. The
hyperparasite population grows following successful transmission to flukes.
This occurs within the cockle and so the probability of successful
transmission (we can assume that the initial hyperparasite infection remains
in a dormant condition prior to infection) will be proportional to
an average number of flukes per cockle. Therefore, the probability
of successful transmission and growth is proportional to $H F_{CO}/C$.
Again, we assume that hyperparasite population dynamics occur on timescales
much less than the other populations, modelled by a small parameter
$\varepsilon\ll 1$:
%(leading to a singularly perturbed system). The differential equation is given by
%\begin{equation}
$$
\varepsilon\frac{dH}{dt}=-cH+\frac{dHF_{CO}}{C}-\frac{g\nu H^{2}}{F_{CO}}.\label{eq:dH}
$$
%\end{equation}
Here $c/d=185$;
%$0<\varepsilon\ll1$ is the scaling factor determining how fast the dynamics occur relative to other populations
and $g$ is a coefficient
%equivalant to $\zeta$ described already as a method of
describing over-dispersion.
%However, again as described in earlier
%sections, the value that we are interested in to measure this will
%be equal to $(g\nu/\varepsilon)$.
This equation has the same form as Eq. (\ref{dH thresh final}),  apart from the fact that we now average the recruitment term over the cockle population.
This should be reasonable given the amount of homogeneity in a given
area.

\subsection{Overall system\label{sec:Numerical populations}}

The system of equations can be summarised as follows:
\begin{equation}
\begin{array}{c}
F_{PL}=\beta P_{I}, \\
\frac{dB}{dt}=k\left(1+p\, Cos\left(\frac{2\pi t}{12}\right)\right)CB+k_{0}B-sB^{2}, \\
\frac{dP_{I}}{dt}=\xi B\,(P_{0}-P_{I})-\gamma P_{I}, \\
\frac{dF_{PM}}{dt}=\kappa(t)F_{PL}-\epsilon F_{PM}, \\
\frac{dC}{dt}=\lambda C-\mu C^{2}-\omega F_{CO}-hCB, \\
\frac{dF_{CO}}{dt}=f\, C\, F_{PM}-\delta F_{CO}-\nu H-\frac{\zeta\omega F_{CO}^{2}}{C}, \\
\varepsilon\frac{dH}{dt}=-cH+\frac{dHF_{CO}}{C}-\frac{g\nu H^{2}}{F_{CO}}. \end{array}\label{eq:S}
\end{equation}
%and we can use a numerical differential equation solver to yield solutions for plotting.
It is not easy to provide an estimate for the parameters
due to their sheer number and the complexity of the system.
However, we wish to be left with reasonable population numbers. We have given hints at some of these numbers above; we will now give a more complete description.

A typical density of cockles can be anything around 1000 per square
meter. However, it was found that
numbers more typical in the south of Ireland were around 100 - 500
(Fermer {\em et al.}, 2010). Since the typical area of a bay could be of the magnitude
of several acres, it is reasonable
to assume that a typical cockle population would be of the order of
millions. We will use numbers of around three million cockles in total.
\emph{S. plana }numbers are much smaller than those of cockles and
we use a fixed population of $P_{0}=20,000$ in the area under consideration.
The number of those plana actually infected with
flukes ranges from about 10\% to 25\% giving numbers of at least $2,000$
to $4,000$ infected plana.

Each cockle can be infected with anything from 100 to 3,000 flukes
or more (Fermer \emph{et al}., 2011), meaning the number of flukes
in cockles can range from about $10^{8}$ to $10^{10}$. These numbers
mean that the total number of hyperparasites could be up to the order
of $10^{12}$ or more. The final population, birds, could be assumed
at peak times to range in the hundreds to possibly thousands, with
essentially absence during warmer Summer months.
\begin{table}[p]
\begin{centering}
\begin{tabular}{|c|c|}
\hline
Parameter & Value\tabularnewline
\hline
\hline
$\beta$ & $10$\tabularnewline
\hline
$k$ & $8\times10^{-6}$\tabularnewline
\hline
$p$ & $0\le p(t)\le 2$\tabularnewline
\hline
$k_{0}$ & $0.088$\tabularnewline
\hline
$s$ & $0.054$\tabularnewline
\hline
$\xi$ & $8\times10^{-6}$\tabularnewline
\hline
$P_{0}$ & $20,000$\tabularnewline
\hline
$\gamma$ & $0.014$\tabularnewline
\hline
$r$ & $50,000$\tabularnewline
\hline
$\kappa$ & $0\le \kappa(t)\le 1$\tabularnewline
\hline
$\epsilon$ & $0.06$\tabularnewline
\hline
$\lambda$ & $0.15812$\tabularnewline
\hline
$\mu$ & $2\times10^{-8}$\tabularnewline
\hline
$\omega$ & $0.0004$\tabularnewline
\hline
$h$ & $0.0001$\tabularnewline
\hline
$f$ & $3.33\times10^{-7}$\tabularnewline
\hline
$\delta$ & $0.1$\tabularnewline
\hline
$\nu$ & $0.05$\tabularnewline
\hline
$\zeta$ & $8$\tabularnewline
\hline
$\varepsilon$ & $0.001$\tabularnewline
\hline
$c$ & $185d$\tabularnewline
\hline
$d$ & $0.0667$\tabularnewline
\hline
$g$ & $0.03$\tabularnewline
\hline
\end{tabular}
\par\end{centering}
\caption{Parameters used for numerical simulations of system \eqref{eq:S}.
%to illustrate the realistic
%population numbers obtained as outlined in Section \ref{sec:Numerical populations}.
%The actual values for modelling the phenomenon for over-dispersal
%for flukes and hyperparasites were set to $3.2$ and $1.5$ respectively
%in this case. (These are calculated as described above as $\zeta\omega/\varepsilon$
%and $g\nu/\varepsilon$ respectively). In the simulations, due to
%the complex system and multiple dimensions involved, we take a value
%of $\varepsilon=1/1000$. This does not impact negatively on the nature
%of a slow-fast system as $\varepsilon\ll1$. Descriptions of each
%of the parameters can be found in Section \ref{sec:Bird-Shellfish-Parasite-model
\label{tab:Numerical parameters}}
\end{table}

\begin{table}[p]
\begin{centering}
\begin{tabular}{|c|c|}
\hline
Parameter & Value\tabularnewline
\hline
\hline
$\beta$ & $10$\tabularnewline
\hline
$k$ & $10^{-6}$\tabularnewline
\hline
$p$ & $0\le p(t)\le 2$\tabularnewline
\hline
$k_{0}$ & $0.88$\tabularnewline
\hline
$s$ & $0.054$\tabularnewline
\hline
$\xi$ & $10^{-5}$\tabularnewline
\hline
$P_{0}$ & $20,000$\tabularnewline
\hline
$\gamma$ & $0.01$\tabularnewline
\hline
$r$ & $25,000$\tabularnewline
\hline
$\kappa$ & $0\le \kappa(t)\le 1$\tabularnewline
\hline
$\epsilon$ & $1$\tabularnewline
\hline
$\lambda$ & $0.15$\tabularnewline
\hline
$\mu$ & $2\times10^{-8}$\tabularnewline
\hline
$\omega$ & $0.00025$\tabularnewline
\hline
$h$ & $0.00009$\tabularnewline
\hline
$f$ & $1.25\times10^{-7}$\tabularnewline
\hline
$\delta$ & $0.0001$\tabularnewline
\hline
$\nu$ & $0.0256$\tabularnewline
\hline
$\zeta$ & $0.00025$\tabularnewline
\hline
$\varepsilon$ & $0.0001$\tabularnewline
\hline
$c$ & $185d$\tabularnewline
\hline
$d$ & $0.0014$\tabularnewline
\hline
$g$ & $1.3$\tabularnewline
\hline
\end{tabular}
\par\end{centering}
\caption{Parameters used for numerical simulations of system \eqref{eq:S}.
\label{tab:Numerical parameters1}}
\end{table}

\subsection{Numerical results}
Table \ref{tab:Numerical parameters} presents a typical set of parameters
resulting in population numbers within expected ranges for multi-species system
\eqref{eq:S}.
%For this section, we are not concerned with
%showing the translation of the original threshold phenomenon into
%this larger, more complete model. Figure \ref{fig:Population-graphics} shows
In particular, the infected \emph{S. plana }population varies from about $2,000$
to over $2,300$ during a one year cycle.
%The overall population of \emph{S. plana }is fixed at $P_{0}=20,000$.
The cockle population
has a low of about $2.8$ million and a high of about $3.4$ million.
The average numbers of flukes per cockle is around $200$ with some
annual variation, and the corresponding hyperparasite population,
always present due to the fluke population remaining above the threshold,
around $29,000$ per cockle.
The bird population dies out, or at least is very small during summer
months, and peaks at about $900$ during the Winter. The corresponding
negative effect that the birds have on cockles can be seen in the
dramatic decline in cockle numbers at this time of year.
The populations of mature flukes in plana and total numbers of flukes
in cockles are also in line with expectations.

\subsection{Threshold effect}

\begin{figure}[h]
\begin{centering}
\includegraphics[scale=0.8]{%threshold_for_large_model.eps
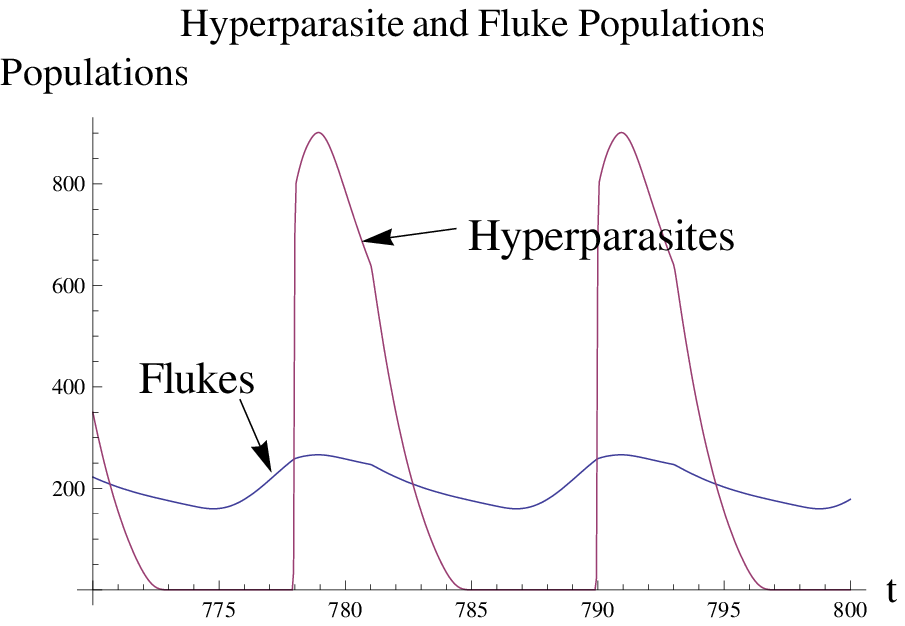} \
\includegraphics[scale=0.8]{%threshold_for_large_model.eps
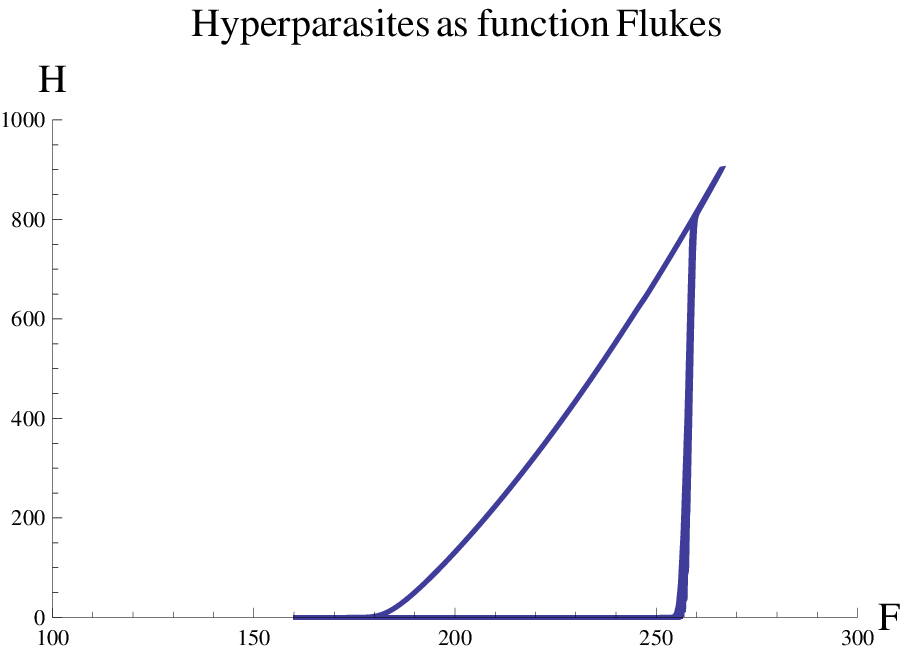}
\par\end{centering}
\caption{Threshold effect evident for system \eqref{eq:S}
with parameter values as given in Table \ref{tab:Numerical parameters1}.
%except for $d=0.00009786$ and $f=1/31,000,000$.
Left: Time traces of the fluke and hyperparasite populations.
Right: Projection of the periodic solution on the fluke-hyperparasite plane.
The hyperparasite can be
seen to be absent from the system during periods in which the average
number of flukes per cockle drops below 185, that is
the threshold type dynamics  similar to that demonstrated by system \eqref{dF thresh final}, \eqref{dH thresh final}, cf. Figure \ref{fig:Seasonal plot}.
\label{fig:Threshold large model}}
\end{figure}

%The main goal of introducing the added populations in this chapter
%was to add to the reality of the observed threshold phenomenon as
%described in the chapter above. We have already shown in the previous
%section that the new and extended model is capable of providing meaningful
%solutions and realistic population numbers. Next,
Figure \ref{fig:Threshold large model} shows a possibility of the translation of the threshold phenomenon
from Eqs. \eqref{dF thresh final}, \eqref{dH thresh final} to the more complex system \eqref{eq:S}.
The hyperparasite population is not present
for part of the year when the fluke population is lower than the threshold value 185.
Once this is reached (or at least, once the exchange of stability
occurs) the hyperparasite population grows rapidly, leading to a momentary
decline in flukes, before an increase in both populations. There are
probably effects of feedback to the system at play here also. The
fluke population grows through immigration which is dependent on birds
which themselves are dependent on cockles. After this point, there
is a steady decline until the fluke population decreases below the
threshold and the hyperparasite population goes to zero until the
new season of fluke immigration begins.
%this larger, more complete model.
%that the model does indeed incorporate the threshold behaviour of
%the fluke on the hyperparasite.
%We are not concerned with providing any exact population numbers (apart, that is, from the threshold
%population of 185 flukes per cockle on average) or anything derived
%from fieldwork results.
%All we wish to show, due to the large number
%of parameter combinations, is that we can artificially perturb the
%system through parameter variation to yield the same dynamics of the
%threshold effect as outlined in the previous chapter.
The parameters used are given in Table \ref{tab:Numerical parameters1}.
%were as those given in Table \ref{tab:Numerical parameters}
%with two changes: the positive effect that flukes have on the
%parasite population $d$ was decreased to $0.00009786$; he growth
%rate of flukes in cockles $f$ was decreased to $3.226\times10^{-8}$.
%

\subsection{The role of the hyperparasite \emph{U. legeri} in the ecosystem}

\begin{figure}[h]
\begin{centering}
\includegraphics{%cockles_during_change.eps
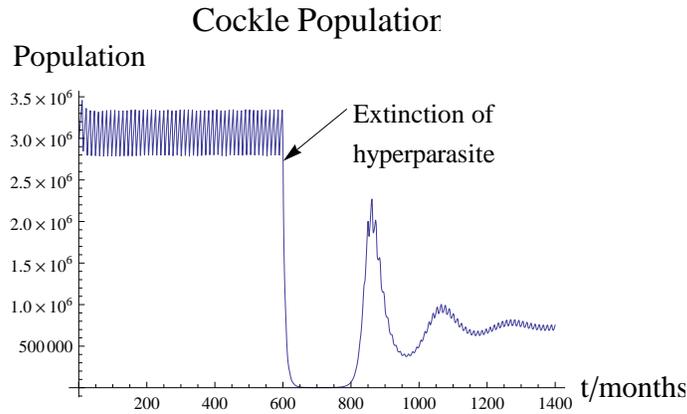}
\par\end{centering}
\caption{Cockle population in system \eqref{eq:S} with parameter values from Table \ref{tab:Numerical parameters1}. After $600$ months, the hyperparasite's death rate is changed to
a greatly increased level leading to their extinction. The effect
this has on cockles is shown here. A stable periodic solution of over
3 million cockles is reached before the hyperparasite is artificially
removed from the system whence there is a dramatic temporary extinction
or absence of cockles from the area for almost 200 months. The effects
that this could have on an ecosystem may be permanently damaging.
However, as the model does not allow for
the cockle population to remain completely extinct, after a period
of renewal, the cockle population settles to a new periodic pattern varying
around the value of $700,000$ cockles. Similar effects are seen in the other
populations indicating the importance that the hyperparasite has on
the system despite its minute size. This effect is indicative that
the hyperparasite is a keystone species.\label{fig:sudden chage}}
\end{figure}

%The equations governing the overall model incorporating all of the
%species that have an impact on each other is given by the system (\ref{eq: Summary of eqs}).
%It has already been shown through numerical simulations carried out
%on this system and the parameter set given in Table \ref{tab:Numerical parameters},
%that the model is well derived and yields accurate and realistic results.
The structure of the interactions of species in Figure 1, and simulations of system \eqref{eq:S},
suggest that the hyperparasite may play a very important role in the ecosystem.
That is, it impacts on the entire system in a much greater way than
its relative size would suggest. In order to provide some verification of this hypothesis, we test a scenario
 where the hyperparasite is effectively
removed entirely from the system at some moment in time, and examine the impact that this
has on the other populations.
The method employed is, using the parameter
values as given in Table \ref{tab:Numerical parameters}, to solve
system \eqref{eq:S} for the time $t=600$ months to ensure that the system reaches
the stable periodic regime. From the time $t=600$ up to the time $t=1400$, the
value of the parameter $c$ is greatly increased. The actual value was
increased from $185\, d$ to $30,000\, d$. This ensures that the
hyperparasite is present only in extremely small numbers and is essentially
wiped out from the system after the moment $t=600$. Figure \ref{fig:sudden chage}
shows the effect
that this change has on the cockle population. It can be
seen that the change is quite dramatic. With the parasite present,
the cockle population varies around $3$ million, depending on time
of year. In the absence of the hyperparasite, this figure drops to
some $700,000$ cockles, a drop of over $70\%$.
The effects on birds, infected plana and other populations (not
shown) are similar. In fact, the bird and infected
plana populations drop by almost $80\%$ after the hyperparasite extinction.

The model \eqref{eq:S} does not allow for the populations to become zero.
Future work may look at a model where it is possible to better quantify the role of the hyperparasite
via a bifurcation scenario.

\section{Conclusions}

We have developed a preliminary model that describes the observed
threshold phenomenon in the cockle system. The interaction of this
fluke-hyperparasite system appears to be of considerable interest. Gam {\em et al.}
(2008) in a study of a coastal lagoon in Morocco found a higher prevalence
of this fluke in older cockles in the inner lagoon compared to the
outer lagoon. One suggestion for the difference in prevalence was
the higher rate of hyperparasitism in the outer lagoon, with $90\%$
of the trematodes being hyperparasited. It was suggested that this
hyperparasite infects and kills \emph{M. minutus} and therefore may control
infrapopulations of this trematode (James {\em et al.} 1977).

Through the use of a relatively standard host-parasite model based
on those developed over the past number of decades, we introduced
the assumption that the hyperparasite dynamics occur in the system
on timescales much faster than that of the fluke. Again this is a
classical slow-fast system which arises in engineering,
biology and physics. The introduction of the small parameter in the
parasite equation gives rise to the threshold, below which no parasites
are observed. In fact, due to the delayed loss of stability, two thresholds
are actually apparent. With the introduction of a mathematical threshold,
we examined the scenario whereby switching of the fluke immigration
on and off occurs. This was described to be a realistic occurrence
due to the periodic nature of fluke maturation in \emph{S. plana}.
It should be noted that it has been observed that although fluke populations
can develop in young cockles, parasites do not appear until the cockle
is over one year old and there also appears to be the building up
of a type of immunity towards flukes in later life. Thus there are
varying factors in the immigration rate that may need consideration
in a future work. We examined immigration effects and, through a simplification
of a switch over a sinusoidal type immigration, we showed that it
was possible to derive algebraic relations, which accurately define the periodic solution in the limit where the ratio of the fluke and hyperparasite time scales tends to zero. In particular, we observe
both delayed and immediate stability exchange in the hyperparasite
population and well-defined indicator times and populations that characterise
the periodic dynamics and the thresholds. Assuming parameter values, we used the equations
relating the unknowns to derive explicitly the moment of the year
when hyperparasite populations start to grow. From this it was possible
also to find other points such as the maximum and minimum fluke populations,
the threshold fluke population and corresponding times.

Finally, we have considered a seasonally driven multi-species model of an ecosystem shown in Figure 1,
which involves bird, and shellfish species typical of a bay or estuarine area, and
accounts for typical stages of the parasites' life cycles.
We have demonstrated that the persistence threshold effect can be transferred
from the cockle-hyperparasite subsystem to the entire ecosystem in this model.
We have also provided evidence that the hyperparasite, despite its minute size,
can be an important regulator of abundance of co-dependent species in the ecosystem.

\section*{Acknowledgments}
This work was supported by the European Regional Development Fund (ERDF) through the Ireland Wales Programme (INTERREG 4A). DR acknowledges the support of the Russian Foundation for basic Research through grant 10-01-93112.


\medskip

\section*{References}
\begin{thebibliography}{100}

\bibitem{}
R M Anderson and R M May 1978
Regulation and stability of host-parasite population interactions: I. Regulatory processes
{\em Journal of Animal Ecology} {\bf 47} 219-247

\bibitem{}
J J Beukema and R Dekker 2005
Decline of recruitment success in cockles and other bivalves in the Wadden Sea: possible role of climate change, predation on postlarvae and fisheries
{\em Mar Ecol Prog Ser} {\bf 287} 149-167

\bibitem{}
 E A Bowers {\em et al.} 1996
The metacercariae of sibling species of Meiogymnophallus, including M. rebecqui comb. nov. (Digenea: Gymnophallidae), and their effects on closely related Cerastoderma host species (Mollusca: Bivalvia)
 {\em Parasitology Research} {\bf 82} 6 505-510

\bibitem{}
V F Butuzov {\em et al.} 2004
Singularly perturbed problems in case of exchange of stabilities
{\em Journal of Mathematical Sciences} {\bf 121} 1973-2079

\bibitem{}
X de Montaudouin, I Kisielewski, G Bachelet  and C Desclaux 2000
A census of macroparasites in an intertidal bivalve community, Arcachon Bay, France
{\em Oceanologica Acta} {\bf 23} 453-468

\bibitem{}
 X de Montaudouin {\em et al.} 2009
 Digenean trematode species in the cockle Cerastoderma edule: identification key and distribution along the north-eastern Atlantic shoreline
 {\em Journal of the Marine Biological Association of the United Kingdom} {\bf 89} 543-556

\bibitem{}
A Deredec and F Courchamp 2003
Extinction thresholds in host-parasite dynamics
{\em Ann Zool Fennici} {\bf 40} 115-130

\bibitem{}
E S Didier, M E Stovall, L C Green, P J Brindley, K Sestak and P J Didier 2004 Epidemiology of microsporidiosis: sources and modes of transmission {\em Veterinary Parasitology} {\bf 126} 145-166

\bibitem{}
J P Ducrotoy {\em et al.} 1991
A comparison of the population dynamics of the cockle (Cerastoderma edule, L.) in North-Western Europe
{\em Estuaries and Coasts: Spatial and Temporal Intercomparisons}
M Elliott and J-P Ducrotoy (eds) (Olsen \& Olsen) 173-184

\bibitem{}
G W Esch, M A Barger and K J Fellis 2002
The transmission of digenetic trematodes: style, elegance, complexity
{\em Integr Comp Biol} {\bf 42} 304–312

\bibitem{}
M Etchechoury and C Muravchik 2003
Nonstandard singular perturbation systems and higher index differential-algebraic systems
{\em Applied Mathematics and Computation} {\bf 134} 323-344

\bibitem{}
J Fermer 2009
Parasitological investigation of softsediment bivalve, with particular reference to the digenean trematode {\em Meiogymnophallus minutus} Ph D Dissertation, University College Cork, Ireland, October 2009, 143 pages


\bibitem{}
J Fermer {\em et al.} 2010
Temporal variation of Meiogymnophallus minutus infections in the first and second intermediate host
{\em Journal of Helminthology} {\bf 84} 362–368

\bibitem{}
J Fermer {\em et al.} 2011
Manipulation of Cerastoderma edule burrowing ability by Meiogymnophallus minutus metacercariae
{\em Journal of the Marine Biological Association of the United Kingdom} {\bf 91} 907–911

\bibitem{}
M Gam, H Bazairi, K T Jensen and X de Montaudouin 2008
Metazoan parasites in an intermediate host population near its southern border: the common cockle (Cerastoderma
edule) and its trematodes in a Moroccan coastal lagoon (Merja Zerga)
{\em Journal of the Marine Biological Association of the United Kingdom} {\bf 88} 357–364

\bibitem{}
 M Gam, X  de Montaudouin and H Bazairi 2010
Population dynamics and secondary production of the cockle Cerastoderma edule: A comparison between Merja Zerga (Moroccan Atlantic coast) and Arcachon Bay (French Atlantic coast)
 {\em Journal of Sea Research} {\bf 63} 191-201

\bibitem{}
W M Getz and J Pickering 1983
Epidemic models: thresholds and population regulation
{\em Am Nat} {\bf 121} 6 892-898

\bibitem{}
 R F Hechinger {\em et al.} 2007
Can parasites be indicators of free-living diversity? Relationships between species richness and the abundance of larval trematodes and of local benthos and fishes
 {Oecologia} {\bf 151} 82-92

\bibitem{}
J A P Heesterbeek and M G Roberts 1995
Mathematical models for microparasites of wildlife
{\em Ecology of Infectious Diseases in Natural Populations}
B T Grenfell and A P Dobson (eds) (Cambridge: Cambridge University Press) 91-122

\bibitem{}
 R D Holt {\em et al.} 2003
Impacts of environmental variability in open populations and communities: "inflation" in sink environments
 {\em Theoretical population biology} {\bf 64} 315-330

\bibitem{}
 P J Hudson {\em et al.} 2006
 Is a healthy ecosystem one that is rich in parasites?
 {\em Trends in Ecology and Evolution} {\bf 21} 7 381-385

\bibitem{}
B L James, A Sannia and E A Bowers 1977
Parasites of birds and shellfish
{\em Problems of a Small Estuary} A Nelson-Smith and E M Bridges (eds)  (Swansea: Proc Burry Inlet Symposium
{\bf 1}) 1-16

\bibitem{}
P T J Johnson, A Dobson, K D Lafferty, D Marcogliese, J Memmott, S Orlofske, R Poulin and D W Thieltges 2010 When parasites become prey: ecological and epidemiological significance
    {\em Trends in Ecology \& Evolution} {\bf 25} 362-371

\bibitem{}
L Kalachev, T C Kelly, M J O'Callaghan, A V Pokrovskii and A A Pokrovskiy 2010
Analysis of threshold-type behaviour in mathematical models of the intrusion of a novel macroparasite in a host colony {\em Math Med Biol} {\bf 28} 4 287-333

\bibitem{}
 A M Kuris {\em et al.} 2008
 Ecosystem energetic implications of parasite and free-living biomass in three estuaries
 {\em Nature} {\bf 454} 515-518

\bibitem{}
G Lauckner {\em et al.} 1983
Diseases of mollusca: Bivalvia
{\em Diseases of Marine Animals Vol. 2 Introduction, Bivalvia to Scaphopoda}
O Kinne (ed) (Hamburg: Biologische Anstalt Helgoland) 477-961

\bibitem{}
 K N Mouritsen and R Poulin 2002
 Parasitism, community structure and biodiversity in intertidal ecosystems
 {\em Parasitology} {\bf 124} S101-S117

\bibitem{}
M Nizette, D Rachinskii, A Vladimirov and M Wolfrum 2006
	Pulse interaction via gain and loss dynamics in passive mode locking
{\em Physica D: Nonlinear Phenomena} {\bf 218} 1 95-104

 \bibitem{}
 T R Raffel {\em et al.} 2008
 Parasites as predators: unifying natural enemy ecology
 {\em Trends in Ecology and Evolution} {\bf 23} 11 610-618

 \bibitem{}
 K Rohde 1994
 Niche restriction in parasites: proximate and ultimate causes
 {\em Parasitology} {\bf  109} S69-S84

\bibitem{}
 F Russell-Pinto 1990
 Differences in infestation intensity and prevalence of hinge and mantle margin Meiogymnophallus minutus Metacercariae (Gymnophallidae) in Cerastoderma edule (Bivalvia): Possible species coexistence in Ria de Aveiro
 {\em J Parasitology} {\bf 76} 5 653-659

\bibitem{}
 F Russell-Pinto {\em et al.} 2006
 Digenean Larvae parasitizing Cerastoderma Edule (Bivalvia) and Nassarius Reticulatus (Gastropoda) from Ria de Aveiro, Portugal
 {\em J Parasitology} {\bf 92} 2 319-332

\bibitem{}
D W Thieltges and K Reise 2006
Metazoan parasites in intertidal cockles Cerastoderma edule from the northern Wadden Sea
{\em Journal of Sea Research} {\bf 56} 284-293

 \bibitem{}
 E R Troemel {\em et al.} 2008
Microsporidia Are Natural Intracellular Parasites of the Nematode Caenorhabditis elegans
 {\em PLoS Biol} {\bf 6} 12 e309

 \bibitem{}
 V Veliov 1997
A generalization of the Tikhonov Theorem for singularly perturbed differential inclusions
 {\em Journal of dynamical and control systems} {\bf 3} 291-319

\bibitem{}
A M Wegeberg and K T Jensen 1999
Reduced survivorship of Himasthla (Trematoda, Digenea)-infected cockles (Cerastoderma edule) exposed to oxygen depletion
{\em Journal of Sea Research} {\bf 42} 4 325-331

\bibitem{}
 A M Wegeberg and K T Jensen 2003
 In situ growth of juvenile cockles, Cerastoderma edule, experimentally infected with larval trematodes (Himasthla interrupta)
 {\em Journal of Sea Research} {\bf 50} 1 37-43




\end{thebibliography}
\end{document}